\documentclass[aps,pra,twocolumn,superscriptaddress]{revtex4-1}
\usepackage{graphicx}
\usepackage{amssymb}
\usepackage{amsmath}
\usepackage{bm}
\usepackage{xcolor}
 \usepackage{verbatim}

\begin{document}

\title{Suppression of Spontaneous Defect Formation in Inhomogeneous Bose Gases}

\author{Myeonghyeon Kim}
\affiliation{Department of Physics and Astronomy, Seoul National University, Seoul 08826, Korea}
\affiliation{Center for Correlated Electron Systems, Institute for Basic Science, Seoul 08826, Korea}

\author{Tenzin Rabga}
\affiliation{Center for Correlated Electron Systems, Institute for Basic Science, Seoul 08826, Korea}

\author{Yangheon Lee}
\affiliation{Department of Physics and Astronomy, Seoul National University, Seoul 08826, Korea}
\affiliation{Center for Correlated Electron Systems, Institute for Basic Science, Seoul 08826, Korea}

\author{Junhong Goo}
\affiliation{Department of Physics and Astronomy, Seoul National University, Seoul 08826, Korea}

\author{Dalmin Bae}
\affiliation{Department of Physics and Astronomy, Seoul National University, Seoul 08826, Korea}
\affiliation{Center for Correlated Electron Systems, Institute for Basic Science, Seoul 08826, Korea}

\author{Y. Shin}
\email{yishin@snu.ac.kr}
\affiliation{Department of Physics and Astronomy, Seoul National University, Seoul 08826, Korea}
\affiliation{Center for Correlated Electron Systems, Institute for Basic Science, Seoul 08826, Korea}
\affiliation{Institute of Applied Physics, Seoul National University, Seoul 08826, Korea}

%\date{\today}

\begin{abstract}

In phase transition dynamics involving symmetry breaking, topological defects can be spontaneously created but it is suppressed in a spatially inhomogeneous system due to the spreading of the ordered phase information. We demonstrate the defect suppression effect in a trapped atomic Bose gas which is quenched into a superfluid phase. The spatial distribution of created defects is measured for various quench times and it is shown that for slower quenches, the spontaneous defect production is relatively more suppressed in the sample's outer region with higher atomic density gradient. The power-law scaling of the local defect density with the quench time is enhanced in the outer region, which is consistent with the Kibble--Zurek mechanism including the causality effect due to the spatial inhomogeneity of the system. This work opens an avenue in the study of nonequilibrium phase transition dynamics using the defect position information.

\end{abstract}

\maketitle

Topological defect creation in the dynamics of phase transition is a remarkable manifestation of the underlying spontaneous symmetry breaking. When a system is cooled to an ordered phase, spatial domains with randomly broken symmetries develop and their merging may lead to formation of topological defects that survive the ordered phase of the system~\cite{Kibble76}. Such spontaneous defect formation is ubiquitous in nonequilibrium phase transition dynamics and has been discussed in various fields such as cosmology~\cite{Kibble76,Kibble80}, condensed matter physics~\cite{Zurek85,Bauerle96,Ruutu96,Chuang91,Carmi00,Sadler06,Weiler08} and recently, quantum computing~\cite{Gardas18,Keesling19,Weinberg20}. A general description of the defect formation dynamics is provided by the Kibble--Zurek mechanism (KZM)~\cite{Zurek96,Dziarmaga10,delCampo14}, where based on the breakdown of the system's adiabatic evolution due to critical slowing down~\cite{Hohenberg77} and the scaling properties of the system near the critical point, the defect density is predicted to exhibit a power-law dependence on the time scale in which the phase transition is crossed. The universality of the KZM has been experimentally tested with many different systems~\cite{Chuang91,Monaco02,Weiler08,Pyka13,Ulm13,Ejtemaee13, Lamporesi13,Navon15,Donadello16,Ko19,Goo21,Goo22}.

In a spatially inhomogeneous system, the defect formation dynamics can be qualitatively changed. Since the local critical temperature varies spatially due to the inhomogeneity, different parts of the system undergo the phase transition at different times during a quench, so that a phase transition front is formed and propagates in the system. If the transition front moves sufficiently slow, the ordered phase behind the front would influence the symmetry breaking at the critical front, consequently suppressing the defect formation~\cite{Kibble97}. Therefore, the defect creation probability is determined not only by the cooling rate but also by the propagation speed of the transition front. The causal independence of distant regions is essential for spontaneous defect formation but is undermined by the spatial inhomogeneity of the system.

To incorporate the causality effect, an extension of the KZM was proposed by imposing an additional defect creation criterion that the propagation speed of the phase transition front should exceed the local phase information spreading speed~\cite{Dziarmaga99,delCampo11,delCampo13}. This model, referred to as inhomogeneous KZM (IKZM), presents two key predictions: (1) appearance of defect-free regions breaking the causal indepedence criterion~\cite{Dziarmaga99} and (2) an enhanced power-law scaling of the total defect number with the quench time~\cite{delCampo11}. In recent experiments with trapped ion chains~\cite{Pyka13,Ulm13, Ejtemaee13} and atomic gases~\cite{Ko19,Goo21,Goo22}, the scaling exponents of the total defect number were measured to be higher than the KZM predictions, supporting the IKZM. However, without comparative measurements for homogeneous systems, the causality-induced defect suppression was not conclusively demonstrated.

In this paper, we investigate the causality effect in spontaneous defect formation by measuring the defect position distributions for various quench times in an inhomogeneous atomic Bose gas. The gas sample is trapped in an external potential to have a spatially varying particle density and upon quenching into a superfluid phase, quantum vortices are spontaneously created in the system. We observe that as the quench time increases, the spatial distribution of defects is localized to the central region of the sample, revealing the defect suppression in the outer region with higher atomic density gradient. It is also observed that the power-law scaling exponent of the local defect density with the quench time is enhanced in the outer region. These observations are qualitatively consistent with the predictions of the IKZM and demonstrate the defect suppression effect by causality in the phase transition dynamics of an inhomogeneous system.

We consider a gas of identical bosonic particles cooled into a superfluid phase, where its temperature is linearly lowered with a quench rate $r_q=-\frac{dT}{dt}$ through the critical temperature $T_c$. The characteristic quench time is given by $\tau_q=T_c/r_q$. In a homogeneous system, according to the KZM, the spatial correlation length at the freeze-out time when the system stops adiabatic following of the quench near the critical region, is given by~\cite{delCampo14}
\begin{equation}
\hat{\xi}= \xi_c  \Big(\frac{\tau_q}{\tau_c}\Big)^{\frac{\nu}{1+\nu z}},
\end{equation}
where $\xi_c$ and $\tau_c$ are the system-specific length and time scales, respectively, and $\nu$ and $z$ are the static and dynamic critical exponents, respectively, which are determined by the universality class of the phase transition. For the superfluid phase transition, $\nu \approx 2/3$~\cite{Donner07} and $z\approx 3/2$~\cite{Navon15}. Since the average size of ordered-phase domains is governed by the length scale of $\hat{\xi}$, the defect density is estimated by $n_d\sim \hat{\xi}^{-(D-d)}\propto \tau_q^{-(D-d)\frac{\nu}{1+\nu z}}$ with $D$ $(d)$ being the dimensionality of the system (defect), exhibiting a power-law dependence on the quench time $\tau_q$.

\begin{figure}
	\includegraphics[width=3.4in]{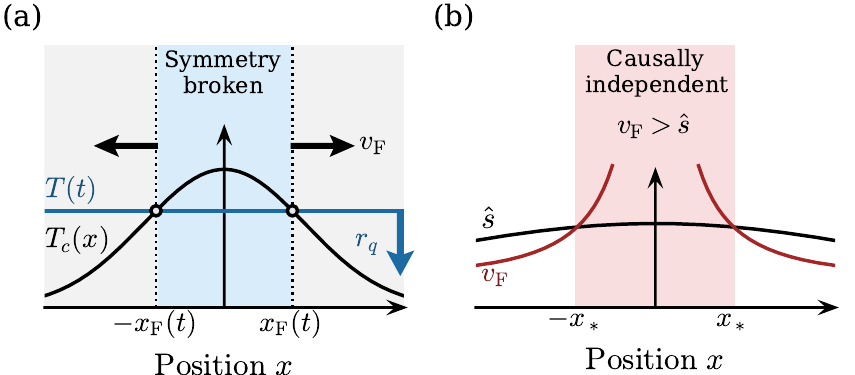}
		\caption{Causality effect in the phase transition dynamics of an inhomogeneous system. (a) The critical temperature $T_c$ varies spatially in an inhomogeneous system, and under a thermal quench with a cooling rate $r_q=-\frac{dT}{dt}$ the front of the ordered phase propagates with a speed $v_\textrm{F}=\frac{d x_\textrm{F}}{dt}$. (b) When $v_\textrm{F}$ exceeds the propagation speed of phase information, $\hat{s}$, the local ordered phases are causally independent, resulting in defect formation. When $v_\textrm{F}$ is less than $\hat{s}$, the broken symmetry is determined by the ordered phase of neighboring regions and the defect formation is suppressed.}
\end{figure}

In the inhomogeneous case, e.g, when the atomic gas is confined to an external potential $V(x)$, the local critical temperature $T_c(x)$ varies spatially and the phase transition occurs at the position $x_\textrm{F}$ with $T(t)=T_c(x_\textrm{F})$ at time $t$. For a given cooling rate $r_q$, the phase transition front propagates through the system with a speed given by [Fig.~1(a)] 
\begin{equation}
v_\textrm{F}(x)=\Big|\frac{dT_c}{dx}\Big|^{-1} r_q.
\end{equation}
On the other hand, the local speed $\hat{s}$ of the phase information propagation is estimated by~\cite{Kibble97,Dziarmaga99} 
\begin{equation}
\hat{s}(x)\sim \frac{\hat{\xi}}{\hat{\tau}}=\frac{\xi_c}{\tau_c} \Big(\frac{\tau_q}{\tau_c}\Big)^{\frac{\nu(1-z)}{1+\nu z}}=\frac{\xi_c}{\tau_c^{\gamma}} \Big(\frac{T_c}{r_q}\Big)^{\gamma-1}, 
\end{equation}
where $\hat{\tau}\sim \hat{\xi}^{z}$ is the system's relaxation time at the freeze-out time and $\gamma=\frac{1+\nu}{1+\nu z}\approx 5/6$.
Then, the causal independence condition, $v_\textrm{F}>\hat{s}$, for spontaneous defect formation is expressed as
\begin{equation}
r_q > r_{q,c}(x)\sim \Big|\xi_c\frac{ d \ln T_c}{d x}\Big|^{1/\gamma}  \frac{T_c}{\tau_c}.
\end{equation}
This means that to produce defects at a given position, there is a threshold quench rate $r_{q,c}$ that depends on the spatial gradient of $T_c$.

More specifically, we consider a system trapped in a power-law potential of the form $V(x)\propto |x|^{n_x}$. The local critical temperature is given by $T_\mathrm{c}(x)\sim T_\mathrm{c0} e^{- |x/\Lambda_x|^{n_x}}$ with $T_\mathrm{c0}$ being the critical temperature at the trap center and $\Lambda_x$ the characteristic spatial extent of the trapped sample. For a Bose gas, up to some numerical factors, $\xi_c$ and $\tau_c$ can be estimated by the de Broglie wavelength $\lambda_\textrm{dB}\propto T_c^{-1/2}$ and the elastic collision time $\tau_\textrm{el}\propto T_c^{-2}$ at the critical point, respectively~\cite{footnote1}. Taking into account the $T_c$ dependence of  $\xi_c$ and $\tau_c$, from Eq.~(4),
\begin{equation}
r_{q,c}(x)\sim \frac{T_{c0}}{\tau_{c0}} \Big(\frac{n_x \xi_{c0}}{\Lambda_x^{n_x}} \Big)^{1/\gamma} |x|^{\frac{n_x-1}{\gamma}} e^{ -(3-\frac{1}{2\gamma})|\frac{x}{\Lambda_x}|^{n_x} }, 
\end{equation}
where the 0 in the subscript indicates the trap center. 
For $n_x>1$ and in a near-center approximation for $|x|\ll \Lambda_x$~\cite{delCampo11}, the causality criterion of Eq.~(4) renders a defect-allowed region as 
\begin{equation}
|x|<x_{*}\sim \Big(\frac{\Lambda_x^{n_x}}{n_x \xi_{c0}}\Big)^{\frac{1}{n_x-1}}\Big(\frac{\tau_{c0}}{T_{c0}}\Big)^{\frac{\gamma}{n_x-1}} r_q^{\frac{\gamma}{n_x-1}},
\end{equation}
predicting a suppression of defect formation in the outer region of the trapped system [Fig.~1(b)]

It is important to note that if the phase transition occurs in a locally independent manner over the whole system, i.e., $\hat{s}=0$, the defect density would be higher in the outer region with low $T_c$ because the local correlation length $
\hat{\xi}(x) \propto T_c^{\frac{3\nu}{1+\nu z}-\frac{1}{2}} r_q^{-\frac{\nu}{1+\nu z}} \approx T_c^{1/2} r_q^{-1/3}$ from Eq.~(1). This is in stark contrast to the prediction from Eq.~(6). Therefore, the spatial distribution of defects can reveal the causality effect in the phase transition dynamics of the inhomogeneous system. In the following experiment, we observe that the defect population is relatively more suppressed in the outer region of the system as the quench rate decreases, which is the main finding of this work.

The experiment was performed with a cold thermal gas of $^{87}$Rb in an optical dipole trap (ODT) with a highly oblate geometry as described in Ref.~\cite{Goo22}. The sample is initially prepared at a temperature of $T_i\approx 440$~nK, containing $\approx3.3\times10^{7}$ atoms, and evaporatively cooled by linearly lowering the ODT depth $U$ from $U_i=1.15U_c$ to $U_f=0.27U_c$ in a variable time $\tau$. Here, $U_c$ is the critical trap depth for Bose--Einstein condensation, determined for an equilibrium sample, where the atom number is measured to be $\approx 3.0 \times10^{7}$. At the end of the quench, the sample temperature is $T_f\approx 50$~nK and the typical atom number is $\approx 1.2\times10^{7}$. In our experiment, the quench time $\tau$ is varied from $2.95~\textrm{s}$ to $11.06~\textrm{s}$. For the entire range of $\tau$, the linear relationship between the sample temperature and the trap depth was confirmed~\cite{Goo21}, ensuring that the sample is sufficiently thermalized during the quench evolution. The cooling rate is $r_q=\frac{T_i-T_f}{\tau}$ and the characteristic quench time is $\tau_{q0}=\frac{T_{c0}}{r_q}\approx 0.96 \tau$ with $T_{c0}\approx 370$~nK. At the critical temperature, $\lambda_\textrm{dB}\approx 0.3~\mu\textrm{m}$ and $\tau_\textrm{el}\approx 1~\textrm{ms}$ for the peak atomic density~\cite{footnote1}.

\begin{figure}[t]
	\includegraphics[width=3.4in]{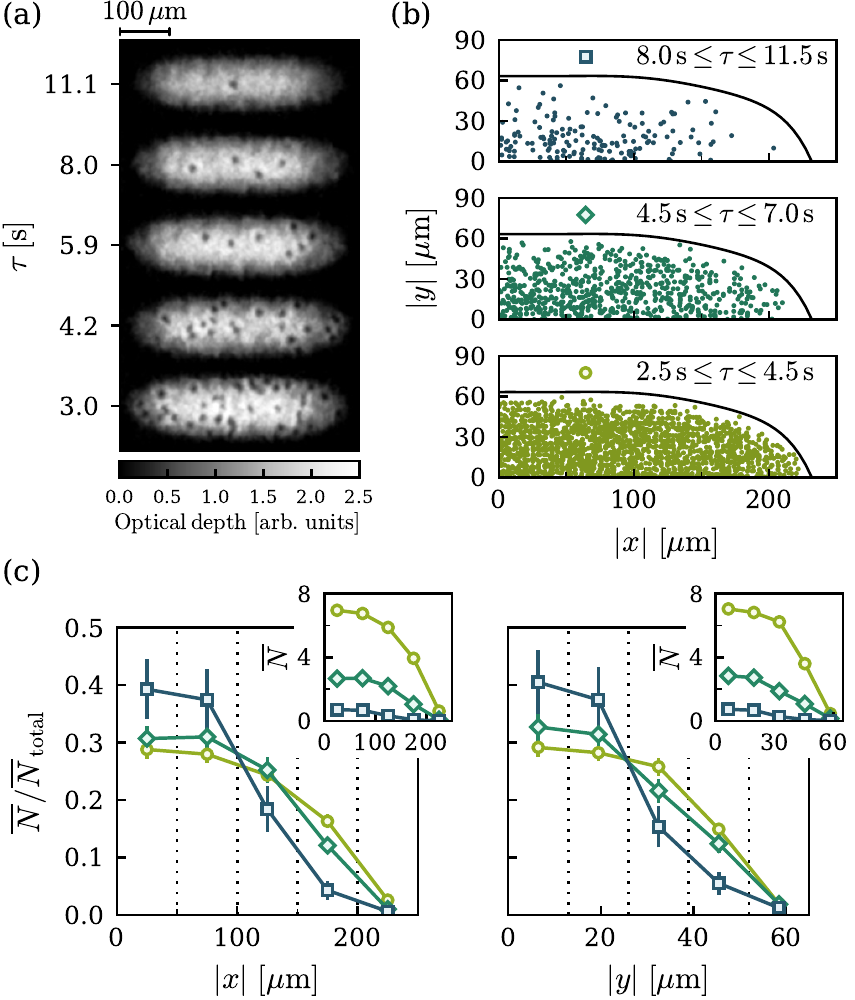}
		\caption{Spatial distribution of defects created in a trapped Bose gas quenched into superfluid phase. (a) Images of Bose gases for various quench times $\tau$. The images were taken after a time of flight and vortices are detected with their density depleted holes in the images. (b) Vortex positions for three different ranges of $\tau$: $8.0{\rm~s}\leq\tau\leq11.5{\rm~s}$, $4.5{\rm~s}\leq\tau\leq7.0{\rm~s}$, and $2.5{\rm~s}\leq\tau\leq4.5{\rm~s}$, where the data were obtained from 90, 80, and 80 realizations of the experiment, respectively. The solid lines indicates the boundary of the condensate determined by the column density threshold at 5\% of the maximum value in the averaged image. (c) Histogram of vortex position in $|x|$ and $|y|$. $\bar{N}$ indicates the mean vortex number in the corresponding spatial bin and $\bar{N}_\textrm{total}$ denotes the mean total number. The markers are same as in (b).}
\end{figure}

After finishing the quench process, a hold time of $\tau_h=1.25~\mathrm{s}$ is applied to facilitate defect formation~\cite{Goo21}. The final condensate fraction is about 80\% and the Thomas--Fermi radii are $R_{x,y,z}\approx (244,65,2.8)~\mu\mathrm{m}$. Due to the tight confinement along the $z$ direction, the sample has quasi-2D geometry and it is energetically favorable to have vortex lines aligned along the short axis. The created vortices are detected by imaging the sample along the $z$ direction after a time-of-flight of 40.4~ms.  

In the experiment, to generate a large area sample, we employ an ODT formed by focusing a clipped Gaussian laser beam~\cite{Lim21}. The transverse trapping potential is determined to be 
\begin{equation}
V(x,y)=V_0\bigg[\Big(\frac{x}{R_x}\Big)^{3.9} +\Big(\frac{y}{R_y}\Big)^{2}\bigg]. 
\end{equation}
For the laser beam clipping, the focal region is elongated and flattened along the beam propagation ($x$) axis, and the trapping potential along the direction is approximately quartic.

In Fig.~2(a), we show absorption images of the samples for various quench times. It is apparent that more vortices are created for faster quenches. We measure the positions of the vortices by hand with aid of a vortex detection algorithm based on a convolutional neural network method~\cite{Metz21,footnote2}. In Fig.~2(b), we display the vortex positions obtained for three different ranges of the quench time and in Fig.~2(c), plot the histograms of the average vortex number $\bar{N}$ (inset) and the relative probability of the vortex appearance, $\bar{N}/\bar{N}_\textrm{total}$, in the different binned regions along the long ($x$) and short ($y$) transverse directions. $\bar{N}_\textrm{total}$ is the average total vortex number of the system. The measurement results clearly show that the vortex creation probability is more suppressed in the outer region for longer quench times, which is consistent with the expectation from the causality effect in a quenched inhomogeneous system.

\begin{figure}[t]
	\includegraphics[width=3.4in]{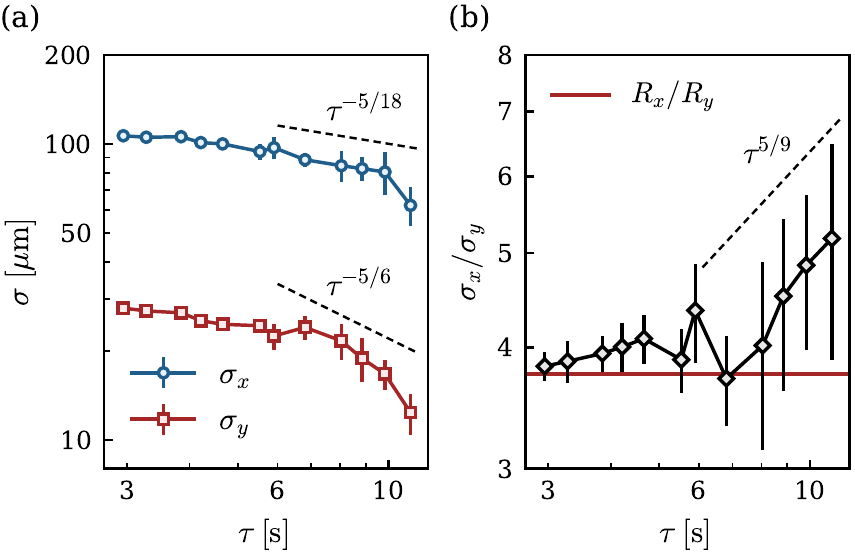}
		\caption{Evolution of the vortex distribution for various quench times. (a) Distribution width $\sigma_{x(y)}$ of vortices as a function of $\tau$. The width $\sigma_{x(y)}$ is determined such that 60\% of vortices are contained within the region with $|x|<\sigma_x ~(|y|<\sigma_y)$. Each data point was obtained from 20 realizations of the same experiment except for $\tau = 11.06{\rm~s}$ with 30 realizations. The uncertainties shown by the error bar were estimated by bootstrap resampling. (b) $\sigma_x/\sigma_y$ as a function of $\tau$. The horizontal solid line indicates the aspect ratio of the sample. The both panels are log-log plots. The dashed lines are drawn for reference, indicating the power-law scaling from Eq.~(6).}
\end{figure}

To characterize the evolution of the vortex distribution, we determine the widths $\sigma_{x}$ and $\sigma_{y}$ of the distribution such that 60\% of the detected vortices are contained within the region with $|x|\leq \sigma_x$ and $|y|\leq \sigma_y$. The uncertainties on the widths are determined by bootstrap resampling repeated 1000 times. Fig.~3(a) shows the evolution of the widths as a function of $\tau$, corroborating their decreasing behavior with the quench time. 

One noticeable observation is that in the slow quench regime with $\tau>6\,\mathrm{s}$, $\sigma_y$ decreases relatively faster than $\sigma_x$, which is also shown in the evolution of their ratio, $\sigma_x/\sigma_y$ in Fig.~3(b). We attribute it to the difference in the power-law dependencies of the trapping potential along the $x$ and $y$ directions. According to Eq.~(6), the dependence of the widths on $\tau$ is expected to be $\sigma_x\propto \tau^{-5/18}$ and $\sigma_y\propto \tau^{-5/6}$ for $n_x=4$ and $n_y=2$, and thus, $\sigma_x/\sigma_y\propto \tau^{5/9}$. In Figs.~3(a) and 3(b), the expected scaling behaviors are indicated by dashed lines and it seems that they qualitatively explain the evolution of the measured widths in the slow quench regime quite well. For $\tau=10~\textrm{s}$, the spatial ranges, $x_*$ and $y_*$, of the defect-allowed region are estimated to be $x_*\approx 115~\mu\textrm{m}$ and $y_*\approx 4~\mu\textrm{m}$ from Eq.~(6) with $\xi_{c0}=\lambda_\textrm{dB}$, $\tau_{c0}=\tau_\textrm{el}$, and $\Lambda_{x,y}=R_{x,y}$, which compare reasonably well to the measured widths.

For fast quenches, however, $\sigma_y$ deviates from the expected trend and particularly, the ratio $\sigma_x/\sigma_y$ seems to settle down at a value close to the sample aspect ratio. To get a complete understanding of the evolution of the measured widths for changing $\tau$, it might be necessary to take into account the effect of vortex motion or diffusion~\cite{Rickinson18} which possibly occurs during the defect formation period in the experiment. 
For our fastest quench, the vortex density in the central region is $n_v\approx 7\times 10^{-4}\,\mu\textrm{m}^{-2}$ and the characteristic drift velocity is estimated to be $v_d=\frac{\hbar}{m}\sqrt{n_v}\approx 20~\mu\textrm{m/s}$.

\begin{figure}[t]
	\includegraphics[width=3.4in]{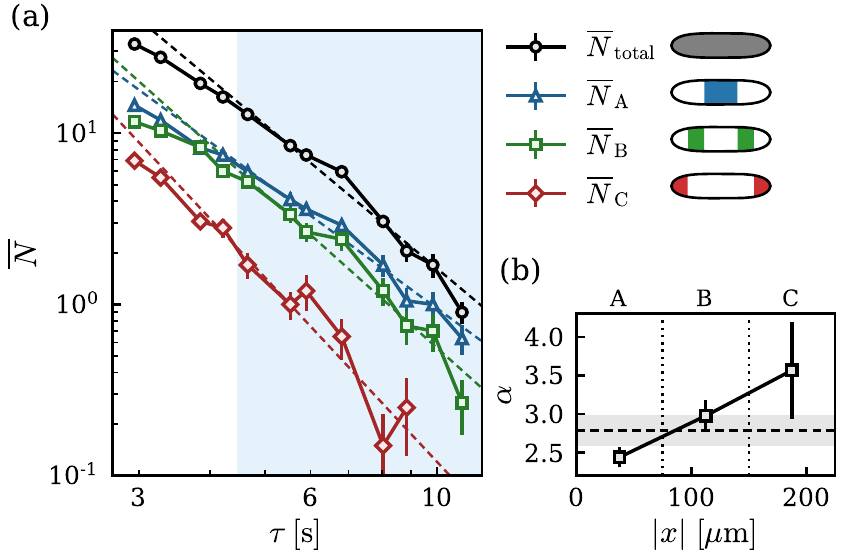}
		\caption{Scaling of the defect number with the quench time. (a) Defect numbers as functions of the quench time $\tau$ for three different regions of the sample along the transverse $x$ direction: the central region (A, blue triangles), the middle region (B, green rectangles), and the outer region (C, red diamonds). The black circles denote the total defect number. The dashed lines are power-law fits to the scaling regime of the corresponding data with $\tau>4.5\,\textrm{s}$ (light blue shaded). The data points below 0.05 are excluded. (b) Power-law exponents for the different regions. The horizontal dashed line and the gray band indicate the value of the scaling exponent for the total defect number and its uncertainty, respectively.}
\end{figure}

From the knowledge of the vortex position, we now investigate the spatial dependence of the scaling exponent of defect density for different regions in the inhomogeneous system. In Fig.~4(a), we plot the mean vortex numbers as functions of the quench time for three different regions of the sample: (A) the center ($|x|$ $<75\,\mu\textrm{m}$), (B) the middle ($75\,\mu\textrm{m}$ $\leq$  $|x|$ $<150\,\mu\textrm{m}$), and (C) the outer regions ($150\,\mu\textrm{m}$ $\leq$ $|x|$). As expected from the causality effect, the vortex number decreases with $\tau$ more rapidly in the outer region. We determine the scaling exponent by fitting a power-law function of $N_{i}=N_{i0} \tau^{-\alpha}$ ($i=$A,B,C) to the data for each sample region and find that the exponent $\alpha$ increases from $\approx 2.4$ for the central region to $\approx 3.6$ for the outer region [Fig.~4(b)]. In the power-law fitting, we used the data in the range $\tau>4.5\,\textrm{s}$ to avoid the saturation effect for fast quenches~\cite{Goo21}.

Although the increment of the local scaling exponent is consistent with the causality effect, the quantitative understanding of the measured exponents is currently lacking. In Ref.~\cite{delCampo11}, the expected total defect number for the inhomogeneous system is estimated as $\bar{N}_\textrm{total}\sim S_*/\hat{\xi}_0^2$ with $S_*\sim x_*y_*$ being the effective area for defect formation, yielding $\bar{N}_\textrm{total}\sim \tau^{-\alpha_\textrm{IKZ}}$ with $\alpha_{\mathrm{IKZ}}=\frac{2\nu}{1+\nu z} +(\frac{1}{n_x-1}+\frac{1}{n_y-1})\gamma\approx16/9$ for our case with $(n_x,n_y)\approx (4,2)$. In Fig.~4(a), the total defect number $\bar{N}_\textrm{total}$ is also displayed and the scaling exponent is measured to be $\alpha_\textrm{total}=2.8(2)$, which is significantly higher than the predicted value. Recently, it was experimentally shown that the scaling behavior of the defect number is also affected by the coarsening dynamics in the early stage of condensate growth~\cite{Goo22,Chesler15}. Since the relative quench depth is shallower in the outer region for lower $T_c$, the early coarsening effect might partially contribute to the observed spatial variations of the local scaling exponent. This possibility is beyond the scope of the IKZM and warrants further investigation in future.

In conclusion, we have investigated the spatial distribution of vortices in a quenched Bose gas with inhomogeneous density distribution and observed that the local suppression of spontaneous defect creation is consistent with the causality effect due to the spatial inhomogeneity of the system. With precise control of the trapping potential, we expect that this work can be extended to a quantitative study of the defect creation probability via statistical analysis of the defect position information for various power-law traps. For example, with a linear trapping potential, the phase information propagation speed could be measured from the onset of defect formation above a certain critical quench rate~\cite{Dziarmaga99}. Furthermore, position correlations between created defects may provide another avenue to understand the details of the defect formation dynamics~\cite{delCampo22}.

\begin{acknowledgments}
This work was supported by the National Research Foundation of Korea (NRF-2018R1A2B3003373, NRF-2019M3E4A1080400) and the Institute for Basic Science in Korea (IBS-R009-D1). 
\end{acknowledgments}


\begin{references}
\bibitem{footnote1}
$\lambda_\textrm{dB}=\frac{h}{(2\pi m k_\textrm{B} T)^{1/2}}$ with $m$ being the particle mass. $\tau_\textrm{el} = (n \sigma \bar{v} )^{-1}$ with $n$ being the particle density, $\sigma=8\pi a^2$ the collisional cross section, and $\bar{v}=4 \sqrt{\frac{k_\textrm{B} T}{\pi m}}$ the mean relative velocity magnitude. $a$ is the $s$-wave scattering length. At the critical point, $n\propto \lambda_\textrm{dB}^{-3}$ and $\tau_\textrm{el}\propto T_c^{-2}$.
\bibitem{footnote2}
Assuming that the relative positions of vortices with respect to the condensate are maintained during the tight-of-flight, we determine the \textit{in situ} vortex position from the measured vortex location in the absorption image.
\end{references}
\end{document}